# Frustration induced noncollinear magnetic order phase in one-dimensional Heisenberg chain with alternating antiferromagnetic and ferromagnetic next nearest neighbor interactions

Jian-Jun Jiang [1,a] , Fei Tang [2], and Cui-Hong Yang[3]

[1]Department of Physics, Sanjiang College, Nanjing 210012, P.R. China

[2] Department of Electronic and Information Engineering, Yangzhou Polytechnic Institute, Yangzhou 225002, P.R. China

[3]Faculty of Mathematics and Physics, Nanjing University of Information Science and Technology, Nanjing 210044, P.R. China

**Abstract.** By using the coupled cluster method, the numerical exact diagonalization method, and the numerical density matrix renormalization group method, we investigated the properties of the one-dimensional Heisenberg chain with alternating antiferromagnetic and ferromagnetic next nearest neighbor interactions. In the classical limit, the ground state is in the collinear Neel state if $\alpha < 1/2$, while for $\alpha > 1/2$ there is an noncollinear canted state. For the quantum case, we found that, although the classical Neel state is absent, the canted state exists if the frustration parameter $\alpha$ exceeds a critical point $\alpha_{c_1}$. The precise critical point $\alpha_{c_1}$ can be determined by using the coupled cluster method and the numerical exact diagonalization method separately. The results of the coupled cluster method and the exact diagonalization method both disclose that the type of phase transition occurring at $\alpha_{c_1}$ changes from a classical second-order transition to a quantum first-order transition due to quantum fluctuation. Although there is another critical point $\alpha_{c_2}$ in a finite system at which the ground state evolves from the canted state to the collinear Neel plus ferromagnetic state, that state is absent because $\alpha_{c_2}$ tends to infinity in the thermodynamic limit.

## 1. Introduction

It is always a fundamental topic to study the properties of the one-dimensional Heisenberg spin chains with frustration in condensed matter physics. The classical ground state of the ordinary one-dimensional spin chain usually possesses the collinear magnetic order state or the noncollinear magnetic order state. A well known prototypical case is the one-dimensional spin-1/2 $J_1 - J_2$ chain. The classical phase diagram of that model exhibits two phases separated by a second-order phase transition at the critical point $\alpha_c = J_2 / J_1 = 0.25$ . For $\alpha < \alpha_c$ , the one-dimensional spin-1/2 $J_1 - J_2$ chain is in the collinear magnetic Neel order state, while for $\alpha > \alpha_c$, that model has noncollinear magnetic spiral order state. However, the collinear magnetic order state of that chain in the quantum case is completely destroyed due to the strong quantum fluctuation in the low dimensionality even in the absence of frustration [1, 2]. When the frustration parameter $\alpha$ exceeds a critical point $\alpha_c \approx 0.241$, the system evolves into a quantum dimerized state, but not the noncollinear magnetic order state [1, 3]. As the classical magnetic order state is rigorously ruled out in the quantum phase diagram due to quantum fluctuation, the question how one can recover the magnetic order state of the one-dimensional spin chain becomes an interesting topic in condensed matter physics [1, 4-8]. There have been some methods to reach this goal. One efficient way is to put the spin chain into a staggered magnetic field which can freeze quantum fluctuation and favors magnetic order [6-8]. The progress foundation of the Neel state of the spin-1 chain in a staggered field is a prototypical case [7]. The other typical method is to add the unfrustrated interactions into the one-dimensional spin chain [1, 4, 5]. For example, it has been approved in reference [5] that the one-dimensional spin-1/2 chain with long-range, non-frustrating interactions possesses collinear magnetic order in some cases.

[a]e-mail: jian_jun_jiang@aliyun.com

In the past researches, often people mainly focused on how to recover the collinear magnetic order state of the quantum spin chain, whereas how to recover the noncollinear magnetic order has been seldom discussed. Here, we investigated the properties of a one-dimensional Heisenberg chain with both frustrated and unfrustrated next nearest neighbor interactions and found that the noncollinear magnetic order can be recovered due to emergence of the unfrustrated next nearest neighbor interaction. As shown in Figure 1, the model Hamiltonian of the one-dimensional Heisenberg chain with alternating antiferromagnetic and ferromagnetic next nearest neighbor interactions is

$$H = J_1 \sum_{i=1}^{N/2} S_{2i} \cdot (S_{2i-1} + S_{2i+1}) + J_2 \sum_{i=1}^{N/2} S_{2i} \cdot S_{2i+2} + J_3 \sum_{i=1}^{N/2} S_{2i-1} \cdot S_{2i+1} \qquad (1)$$

where $S_{2i}$, $S_{2i-1}$ and $S_{2i+1}$ are spin-1/2 operators, $J_1$ is the nearest neighbor antiferromagnetic interaction, $J_2$ is the frustrated antiferromagnetic interaction, and $J_3$ is the unfrustrated ferromagnetic interaction. The number of the unit cells is denoted by $N$/2, and then the total number of sites is $N$. For convenience, the above spin-1/2 $J_1 - J_2 - J_3$ chain can be divided into two sublattices (denoted A and B) corresponding to geometric position difference as shown in Figure 1. Moreover, in what follows we set $J_2 = \alpha J_1 = \alpha$ and $J_3 = -1$. In the case of $J_3 = 0$, the $J_1 - J_2 - J_3$ chain is reduced to the sawtooth chain and the properties of that model has been discussed in reference [9] detailedly. The results of reference [9] disclose that, the collinear Neel order, the noncollinear canted order, and the spiral order in the quantum sawtooth chain are completely destroyed by quantum fluctuation although that model possesses the above three magnetic order in the classical limit. In the present paper, we focus on the effect of frustration $J_2$ on the properties of the sawtooth chain in the presence of unfrustrated ferromagnetic interaction $J_3$ which favors the magnetic order by using the coupled cluster method (CCM), numerical exact diagonalization (ED) method, and numerical density matrix renormalization group (DMRG) method (if necessary).

The paper is organized as follows. In the following section, the details of the application of CCM formalism to the $J_1 - J_2 - J_3$ chain are described. In section 3, the results of the analytical and numerical method are presented. A summary is given in the final section.

## 2. The coupled cluster method applied to the $J_1 - J_2 - J_3$ chain

In recent years, CCM has been very successfully applied to different quantum spin chains [10-26]. One main advantage of CCM consists in its applicability to spin chain with frustration in any dimension. The interested reader can obtain the detailed descriptions of the CCM applied to quantum spin systems in the papers [10, 12, 13]. Here, we only briefly outline the application of CCM to the $J_1 - J_2 - J_3$ chain. The starting point for a CCM calculation is to choose a normalized model state $|\phi\rangle$ and this is often a classical spin state. Classically, the $J_1 - J_2 - J_3$ chain is in the collinear Neel state if $\alpha < \alpha_c = 1/2$ and it possesses the noncollinear canted state as shown in Figure 1 for $\alpha > \alpha_c$. The canted angle $\theta$ is given by $\theta = \cos^{-1}(1/2\alpha)$. Therefore, we chose the Neel state $|\downarrow\uparrow\downarrow\uparrow\cdots\rangle$ for small values of the frustration parameter but canted state characterized by a pitch angle $\theta$ $|\downarrow\nwarrow\downarrow\nearrow\cdots\rangle$ for large $\alpha$ as the model state. Obviously, the canted state is just the Neel state if $\theta = 0$. In the CCM calculation based on canted state, we do not choose the classical pitch angle, but consider the pitch angle as a free parameter and determine it by minimizing the ground state energy with respect to the pitch angle , because the pitch angle may be affected by quantum fluctuation. Then we perform a rotation of the local axes of the spins so that all spins in the model state align along the negative *z*-axis. After this rotation, the CCM parameterization of the ket and bra ground states of model (1) are given by [12, 13]

$$|\psi\rangle = e^{S}|\phi\rangle, \quad S = \sum_{l=1}^{N} \sum_{i_1, i_2, \cdots i_l} S_{i_1, i_2, \cdots i_l} s_{i_1}^{+} s_{i_2}^{+} \cdots s_{i_l}^{+}$$

$$\langle\tilde{\psi}| = \langle\phi|\tilde{S} e^{-S}, \quad \tilde{S} = 1 + \sum_{l=1}^{N} \sum_{i_1, i_2, \cdots i_l} \tilde{S}_{i_1, i_2, \cdots i_l} s_{i_1}^{-} s_{i_2}^{-} \cdots s_{i_l}^{-} \qquad (2)$$

The CCM formalism is exact if we consider all spin configurations in the $S$ correlation operator, but it is usually impossible in practice. A big advantage of the CCM compared to some other methods is the possibility to truncate $S$ in a very systematic and reasonable way. In the present paper, a quite general approximation scheme called LSUB*n* to is used to truncate the expansion of the operators $S$ and $\tilde{S}$ [12, 13]. In the LSUB*n* approximation, only the configurations involving *n* or fewer correlated spins which span a range of no more than *n* contiguous lattice sites are retained. The fundamental configurations retained in the LSUB*n* approximation can be reduced if we choose the collinear Neel state as the reference state because the ground state lies in the subspace $S_{tol}^{z} = \sum_{i=1}^{N} S_i^{z} = 0$ and the Hamiltonian of equation (1) commutes with $S_{tol}^{z}$. For the canted state, one can not reduce the fundamental configurations because it is not an eigenstate of $S_{tol}^{z}$. Moreover, numerical complexity of the CCM based on canted state increases because the determination of the quantum pitch angle requires the iterative minimization of the ground state energy. Therefore, for the Neel model state, we carry out CCM up to the LSUB14 level, whereas we do this only up to the LSUB8 level for the canted state.

To determine the correlation coefficients $S_{i_1, i_2, \mathrm{L} i_l}$ and $\tilde{S}_{i_1, i_2, \cdots i_l}$ contained in the operators $S$ and $\tilde{S}$, one has to solve the following CCM equations [12, 13]

$$\langle\phi| s_{i_1}^{-} s_{i_2}^{-} \cdots s_{i_l}^{-} e^{-S} H e^{S} |\phi\rangle = 0$$

$$\langle\phi| \tilde{S} e^{-S} \left[H, s_{i_1}^{+} s_{i_2}^{+} \cdots s_{i_l}^{+}\right] e^{S} |\phi\rangle = 0 \qquad (3)$$

As the number of the equations grows quickly with the increasing level of approximation, the derivation of the coupled equations for higher orders of approximation is extremely tedious. So we have developed our own programme using Matlab to automate this process according to the method discussed in paper [13].

After the correlation coefficients retained in the LSUB*n* approximation have been obtained, one can use them to calculate the ground state expectation value of some physical observables of the $J_1 - J_2 - J_3$ chain. The first physical quantity that we calculate is the ground state energy from the relation $E_g = \langle \phi | e^{-S} H e^{S} | \phi \rangle$. In order to obtain exact results in the limit $n \to \infty$, the 'raw' LSUB*n* results have to be extrapolated. Although no exact extrapolation rule is known, there is some empirical experience available regarding how the physical quantities might scale with *n*. For the ground state energy, the following formula is used [25]

$$E_g(n)/N = a_0 + a_1 (\frac{1}{n})^2 \tag{4}$$

To check whether the $J_1 - J_2 - J_3$ chain has a long-range order, we also calculate the magnetic order parameter which is given by $M = -\frac{1}{N} \sum_{i=1}^{N} \langle \tilde{\psi} | s_i^z | \psi \rangle$. For the extrapolation of $M$, we use the following two formulas [25, 26]

$$M(n) = b_0 + b_1 n^{-0.5} + b_2 n^{-1.5} \tag{5}$$

$$M(n) = b_0 + b_1 n^{-1} + b_2 n^{-2} \tag{6}$$

or the formula [25, 26]

$$M(n) = b_0 + b_1 n^{-\upsilon} \tag{7}$$

### 3. Results

Since the unfrustrated ferromagnetic interaction $J_3$ favors the collinear Neel order, it is necessary to investigate whether the order exists in the $J_1 - J_2 - J_3$ chain in the absence of $J_2$ firstly. In that case, our extrapolated CCM results for *M* using the scheme of equation (5) and that of equation (6) with *n* = {8, 10, 12, 14} are respectively 0.0026 and 0.0901. Obviously, Similar to the previous research, the value of the extrapolated result of *M* using equation (6) is larger than that using equation (5) [24]. The value of the magnetic order parameter *M* obtained from CCM is so small that one can not judge whether the $J_1 - J_2 - J_3$ chain possesses the collinear Neel order when $\alpha = 0$. To investigate the above problem accurately, we also resort to numerical DMRG method which is a powerful tool to analyse the properties of one-dimensional spin chain calculating the correlation function of the $J_1 - J_2 - J_3$ chain [27]. Corresponding to the two sublattice structure of the $J_1 - J_2 - J_3$ chain shown in Figure 1, two kinds of two-spin correlation functions are defined

$$C_{AA}(|i-N/4|) = \langle\psi|S_{A,i}\cdot S_{A,N/4}|\psi\rangle$$
$$C_{BB}(|i-N/4|) = \langle\psi|S_{B,i}\cdot S_{B,N/4}|\psi\rangle \quad (8)$$

where $|\psi\rangle$ is the ground state. The spin correlation function should converge to a finite value for increasing $|i-N/4|$ and show a power-law behavior if the $J_1-J_2-J_3$ chain has the long-range collinear Neel state [5]. Figure 2 displays the calculation results of correlation function in a system with $N$=612 by using DMRG under open boundary condition. As shown Figure 2, $C_{AA}$ ($C_{BB}$) decreases with the increase of the distance between lattice $i$ and lattice $N/4$ and does not show any convergence behavior. Using second order polynomial fits in $1/(|i-N/4|)$, the infinite size extrapolated values for $C_{AA}$ ($C_{BB}$) can be obtained, as shown in Figure 2. It can be found that $C_{AA}$ ($C_{BB}$) tends to zero in the limit $|i-N/4|\to\infty$. Therefore, as predicted by using the many-body Green' s function theory, the collinear Neel order does not exist in the $J_1-J_2-J_3$ chain [28]. In the presence of frustration $J_2$, the $J_1-J_2-J_3$ chain certainly has not the long-range collinear Neel state since frustration weakens that state. In the following discussion, we call the ground state of the $J_1-J_2-J_3$ chain with $\alpha=0$ the collinear quasi-Neel state.

Next, we discuss how the pitch angle $\theta$ evolves with the increase of frustration $J_2$ by using CCM. In the case of LSUB6 approximation, results for the ground state energy per site $e=Eg/N$ as a function of $\theta$ are shown in Figure 3. According to the structure of the curves displayed in Figure 3, one can divide the full $\alpha$ parameter regime into three parts: $\alpha\le\alpha_{t_1}=0.97$, $\alpha_{t_1}<\alpha<\alpha_{t_2}=1$ and $\alpha\ge\alpha_{t_2}$. As shown in Figure 3, the curves have only one minimum in the first or last parameter regime. In the intermediate parameter regime, two minimums appear. Moreover, in that parameter regime, the location of the global minimum shifts from zero to a finite value at a critical point $\alpha_{c_1}$. The appearance of the two-minimum structure for the ground state energy as a function of $\theta$ indicates that the quantum phase transition occurring at $\alpha_{c_1}$ belongs to a first-order phase transition [17]. The quantum pitch angle obtained from various CCM LSUB$n$ levels is displayed in Figure 4. If one considers the pitch angle as an order parameter, the ground state of the $J_1-J_2-J_3$ chain is in the collinear quasi-Neel state when $\alpha < \alpha_{c_1}$ since $\theta$ equals zero exactly at that parameter regime. And it evolves into the noncollinear canted state characterized by $\theta \neq 0$ if $\alpha$ reaches the critical point. One can also find that at each LSUB*n* level $\theta$ displays a sharp rise at the critical point $\alpha_{c_1}^{LSUBn}$, different from its classical counterparts. Thus at each LSUB*n* level of approximation, the transition from the collinear quasi-Neel state to the noncollinear canted state is first-order. But, as the jump height of $\theta$ at the critical point decreases with the increase of *n*, we can not judge whether the phase transition occurring at $\alpha_{c_1}$ belongs to the first-order transition in the limit $n \to \infty$. The critical point $\alpha_{c_1}$ at each LSUB*n* level drawn from CCM is shown in Figure 5. It can be found that the position of the critical point determined by CCM scales as

$$\alpha_{c_1}(n) = \alpha_{c_1}(\infty) + \frac{c}{n^2} \tag{9}$$

An extrapolation to the limit $n \to \infty$ based on data sets {$\alpha_{c_1}^{LSUB4}$, $\alpha_{c_1}^{LSUB6}$, $\alpha_{c_1}^{LSUB8}$} shows that the critical point is $\alpha_{c_1}^{LSUB\infty} = 0.9171$. To disclose the effect of quantum fluctuation on pitch angle, the classical pitch angle is also displayed in Figure 4. Although the evolution of the quantum angle is similar to that of the classical pitch angle, quantum fluctuation still has significant impact on pitch angle. Besides the sudden jump behavior at the critical point $\alpha_{c_1}$ mentioned above, one can also find the following two facts: (1) The collinear state can extend into the regime $\alpha > 1/2$, where classically it is already unstable. (2) As $\alpha \to \infty$, the quantum pitch approaches its limiting value $\pi/2$ faster than the classical angle does. It is obvious that the spins connected by $J_2$ interactions becomes Neel-ordered and the spins connected by $J_3$ interactions becomes ferromagnetically ordered in the limit $\alpha \to \infty$. We call the ground state of the $J_1 - J_2 - J_3$ chain at that limit the Neel plus ferromagnetic state in the following discussion. According to the above facts, one can reasonably conclude that quantum fluctuation tends to favor collinear state which may exist in the $J_1 - J_2 - J_3$ chain. This phenomenon has also been observed in some other quantum spin chains [17, 25].

Figure 6 shows the ground state energy per site $e = Eg / N$ obtained from CCM and ED. The extrapolated results of CCM based on canted state using equation (4) with the data sets *n* = {4, 6, 8} are also displayed in that Figure. The energies drawn from ED are extrapolated to the thermodynamic limit by using the following formula with $N$=16, 20, 24, 28 and 32 spins [29]

$$f(\alpha, N) = f(\alpha) + c_1 \frac{\exp(-N / c_2)}{N^p} \tag{10}$$

where $p$=2. It is known that the ground state energy density $e$ or its derivative may exhibit a special property at the critical point. It can be seen from Figure 6 that the ground state energy per site $e$ given by CCM converges rapidly with the increase of $n$ and they compare extremely well to those of ED in the whole parameter regime. Both of the results obtained from CCM and ED show a cusp at the critical point $\alpha_{c_1}$, and this is shown in the inset of Figure 6. The appearance of the cusp at the critical point indicates that there is a discontinuity in the first derivative of the energy at that point. Thus, the quantum phase transition taking place at $\alpha_{c_1}$ is first-order. To check that conclusion, the first derivative of the ground state energy of a finite system $dE_g / d\alpha$ obtained from ED is also calculated and plotted in Figure 7. One can observe that a jump indeed exists in that curve at the critical point. Moreover, finite-size effect on the height of the jump is so slight that one can infer reasonably that the first derivative of the ground state energy is discontinuous in the thermodynamic limit. Therefore, just as CCM predicts, the numerical ED results provide us with a huge confirmation that the quantum phase transition occurring at $\alpha_{c_1}$ belongs to a first-order transition. The phase transition in the $J_1 - J_2 - J_3$ chain changing from a classical second-order type to a quantum first-order type is the result of quantum fluctuation. Similar situations have also been seen in other quantum spin systems, such as the spin-1/2 interpolating square-triangle Heisenberg antiferromagnet and the square-lattice Heisenberg antiferromagnet with two kinds of nearest-neighbor bonds [17, 30]. The critical point $\alpha_{c_1}$ obtained from ED is displayed in Figure 8. One can see that the location of the critical point of finite system size $\alpha_{c_1}(N)$ evolves as

$$\alpha_{c_1}(N) = \alpha_{c_1}(\infty) + \frac{c}{N} \tag{11}$$

We used numerical results for $N$=16, 20, 24, 28 and 32 and found that the extrapolation to the thermodynamic limit gives that the critical point is $\alpha_{c_1}(\infty) = 0.9374$. Apparently, the critical point deduced from ED is well consistent with the one obtained from CCM.

In order to investigate whether the $J_1 - J_2 - J_3$ chain possesses the "true" canted order when $\alpha > \alpha_{c_1}$, we also calculated the magnetic order parameter $M$ by using CCM based on canted state. The LSUB*n* results for $M$ with *n* = {4, 6, 8} as well as the extrapolated result using equation (7) are shown in Figure 9. As the $J_1 - J_2 - J_3$ chain does not possess the magnetic long-range order when $\alpha < \alpha_{c_1}$, the magnetic order parameter $M$ at that parameter should be zero. But, we found that it is difficult to obtain the correct result $M = 0$ even by extrapolating the `raw' LSUB*n* data to $n \to \infty$. Thus, the extrapolated results of $M$ are not being displayed in Figure 9 if $\alpha < \alpha_{c_1}$. However, one can still find a sharp rise in $M$ at the critical point $\alpha_{c_1}$ from the results for various LSUB*n* approximations. This finding supports the above conclusion that a first-order phase transition happens at $\alpha_{c_1}$. It can be found that the value of the magnetic order parameter $M$ is finite in the parameter regime $\alpha > \alpha_{c_1}^{LSUB4}$, which means that the magnetic long-range noncollinear canted state exists at that region.

To prove the conclusion given by CCM, we also calculate the total spin of the ground state $S_g$ of the $J_1 - J_2 - J_3$ chain by using ED. Figure 10 displays the results of $S_g$. From that Figure, we find that $S_g$ always equals 0 as $\alpha < \alpha_{c_1}$, which discloses that the collinear quasi-Neel state is always the ground state of the $J_1 - J_2 - J_3$ chain when $\alpha < \alpha_{c_1}$. In the intermediate parameter regime $\alpha_{c_1} < \alpha < \alpha_{c_2}$, $S_g$ takes values between 0 and $N/2$ and it increases with respect to the strength of $\alpha$. At another critical point $\alpha_{c_2}$, the ground state of a finite system evolves into the Neel plus ferromagnetic state characterized by $S_g = N/2$. As the strength of quantum fluctuation which favors collinear order decreases with the increase of the length of the $J_1 - J_2 - J_3$ chain, one can find from Figure 10 that, the location of the critical point $\alpha_{c_1}$ shifts to the left, whereas $\alpha_{c_2}$ shifts to the opposite direction when *N* increases. As a result, the length of the intermediate parameter regime increases with the increase of the system size. Therefore, the $J_1 - J_2 - J_3$ chain does have the magnetic long-range noncollinear canted state with $0 < S_g < N/2$ in the parameter region $\alpha_{c_1} < \alpha < \alpha_{c_2}$.

Searching for the collinear state of the spin systems in the strongly frustrated regime is a challenging and interesting issue in the recent years [31-33]. For the $J_1 - J_2 - J_3$ chain, we also faced the question that quantum fluctuation is strong enough to change the location of the critical point $\alpha_{c_2}$ from infinity in the classical case to a finite value in the quantum case. Since the collinear Neel plus ferromagnetic state is a special case of the canted state with pitch angle $\theta = \pi / 2$, we can first judge whether the phase transition from the noncollinear canted state to the Neel plus ferromagnetic state occurs at a finite $\alpha_{c_2}$ by using the pitch angle $\theta$ obtained from CCM . Figure 11 shows the results of $\theta$ in the case of LSUB6 approximation for large values of $\alpha$ in the range $990 \le \alpha \le 1000$. That Figure shows that, $\theta$ and $\pi / 2$ are not the same value, although they are extremely close to each other at that large parameter regime. Thus, from the results of CCM, one can not find the indication of the appearance of the collinear Neel plus ferromagnetic state in the $J_1 - J_2 - J_3$ chain when $\alpha$ is finite. Next, we also used ED method to discuss the above problem. Since the *z*-component of the total spin $S_{tol}^z$ commutes with Hamiltonian (1), the following relations should be satisfied

$$\begin{aligned} E_1(\alpha, S_{tol}^z = N/4-1, N) &< E_2(\alpha, S_{tol}^z = N/4, N) \qquad \alpha < \alpha_{c_2} \\ E_1(\alpha, S_{tol}^z = N/4-1, N) &= E_2(\alpha, S_{tol}^z = N/4, N) \qquad \alpha \ge \alpha_{c_2} \end{aligned} \tag{12}$$

where $E_1$ and $E_2$ are the energies of the lowest-lying state with $S_{tol}^z = N/4-1$ and $S_{tol}^z = N/4$. Then one can obtain the critical point $\alpha_{c_2}$ by only comparing $E_1$ with $E_2$. The critical point $\alpha_{c_2}$ determined by ED is plotted as a function of the system size *N* in Figure 12. The curve in Figure 12 discloses that $\alpha_{c_2}$ almost increases linearly with the growth of *N*. It means that the value of $\alpha_{c_2}$ inclines to infinity in the thermodynamic limit. Thus, just as predicted by results of CCM, the collinear Neel plus ferromagnetic state is absent in the $J_1 - J_2 - J_3$ chain although quantum fluctuation favors that state.

Till now, we have known that the $J_1 - J_2 - J_3$ chain always has the noncollinear canted state if $\alpha > \alpha_{c_1}$, but the effect of quantum fluctuation on the classical canted state of the $J_1 - J_2 - J_3$ chain has not been clarified. Due to the combined effect of quantum fluctuation and frustration, some physical properties of the noncollinear state, such as two-spin correlation function and local magnetization, often display an incommensurate modulation behavior that can not be understood within the classical picture [34]. Here, we calculate the two-spin correlation function defined as equation (8) using DMRG to investigate whether the above interesting phenomenon exists in the $J_1 - J_2 - J_3$ chain. The results of the two-spin correlation function are shown in Figure 13 for $\alpha = 1$, $\alpha = 1.4$, and $\alpha = 1.8$ and $N = 280$ with open boundary condition. It is evident that the correlation $C_{BB}$ shows an antiferromagnetic behavior while $C_{AA}$ displays a ferromagnetic behavior. One can also clearly observe that, an incommensurate modulation with long-distance periodicity does exist in the correlation $C_{AA}$ in all cases, although the behavior of $C_{BB}$ becomes more and more similar to that of the one-dimensional spin-1/2 chain with the increase of $\alpha$ as expected.

## 4. Conclusions

In this paper, we investigated the properties of the one-dimensional Heisenberg chain with alternating antiferromagnetic and ferromagnetic next nearest neighbor interactions which can be regarded as a variant of the sawtooth chain by using the CCM, the numerical ED and DMRG method. The results disclose that, the long-range collinear Neel order is still absent similar to the sawtooth chain although the $J_1 - J_2 - J_3$ chain contains the unfrustrated ferromagnetic interaction $J_3$ which favors that order. However, different from the sawtooth chain, the $J_1 - J_2 - J_3$ chain possesses the noncollinear canted state with an incommensurate modulation with long-distance periodicity after the frustration $J_2$ exceeds a critical point. The precise location of that point is determined by ED and CCM. We also found that the ground state energy given by CCM agrees excellently with that obtained from ED. This confirms that CCM is a reliable analytical method to investigate the properties of such spin systems with frustration, even in the strongly frustrated regime. The following two interesting topics of spin systems are also discussed in the present paper. One is about the nature of the quantum phase transition occurring at $\alpha_{c_1}$. By analyzing the structure for the ground state energy as a function of the pitch angle near the critical point and the critical behavior of the pitch angle using CCM based on canted state, one can predict that transition is a first-order transition. That conclusion is also checked by the appearance of the cusp in the ground state energy curve at the critical point given by CCM and ED. Moreover, it is fortunately to find that there is a jump in the curve of the first derivative of the ground state energy of a finite system and the jump does exist in the thermodynamic limit. Thus, due to quantum fluctuation, the nature of that transition evolves from a classical second-order type to a quantum first-order type indeed. Determining whether the $J_1 - J_2 - J_3$ chain possesses the collinear Neel plus ferromagnetic state is the other issue that we are interested in. The results of CCM disclose that the pitch angle does not become exactly $\pi/2$ in large $\alpha$ case, which means that above state is absent in the $J_1 - J_2 - J_3$ chain. Although the results of ED indicate that the collinear Neel plus ferromagnetic state appears in the finite size system, the value of the critical point at which the canted state gives way to the collinear Neel plus ferromagnetic state almost increases linearly with the growth of the system size. Thus, the canted state in the $J_1 - J_2 - J_3$ chain persists for all values $\alpha > \alpha_{c_1}$ in the thermodynamic limit.

## Acknowledgments

This work is supported by the Natural Science Foundation of Jiangsu Province under Grant Nos. BK20131428, and the Natural Science Foundation of the Jiangsu Higher Education Institutions under Grant Nos. 13KJD140003.

## Figure captions

Fig.1. The sketch of the classical canted state of the $J_1 - J_2 - J_3$ chain. $\theta$ measures the deviation of the classical spins from the *z* axis.

Fig.2. The spin correlation function $C_{AA}$ $(C_{BB})$ as a function of $|i - N/4|$ when $\alpha = 0$. The dashed line in the inset is the polynomial fit of the form $a_0 + a_1(|i - N/4|)^{-1} + a_2(|i - N/4|)^{-2}$.

Fig.3. The ground state energy per site $e$ versus the canted angle $\theta$ using CCM based on canted state at the LSUB6 level of approximation.

Fig.4. The quantum pitch angle $\theta$ versus $\alpha$.

Fig.5. Finite-size scaling of $\alpha_{c_1}$ given by CCM versus $1/n^2$. The solid line is the fit line.

Fig.6. The ground state energy per site $e$ obtained from CCM and ED. The inset shows the behaviour of $e$ in the vicinity of the critical point $\alpha_{c_1}$.

Fig.7. The first derivative of the ground state energy $dE_g / d\alpha$ as a function of $\alpha$ for *N*=16, *N*=24, and *N*=32.

Fig.8. Finite-size scaling of $\alpha_{c_1}$ given by ED versus $1/N$. The solid line is the fit line.

Fig.9. The magnetic order parameter $M$ versus $\alpha$.

Fig.10. The total spin $S_g$ as a function of $\alpha$ for *N*=16, *N*=24, and *N*=32.

Fig.11. The quantum pitch angle $\theta$ obtained within the CCM LSUB6 approximation in the large $\alpha$ regime.

Fig.12. Finite-size scaling of $\alpha_{c_2}$ given by ED versus $1/N$. The solid line is the fit line.

Fig.13. The spin correlation function $C_{AA}$ $(C_{BB})$ as a function of $|i - N/4|$ in a system with *N*=256.

**Figure 1**

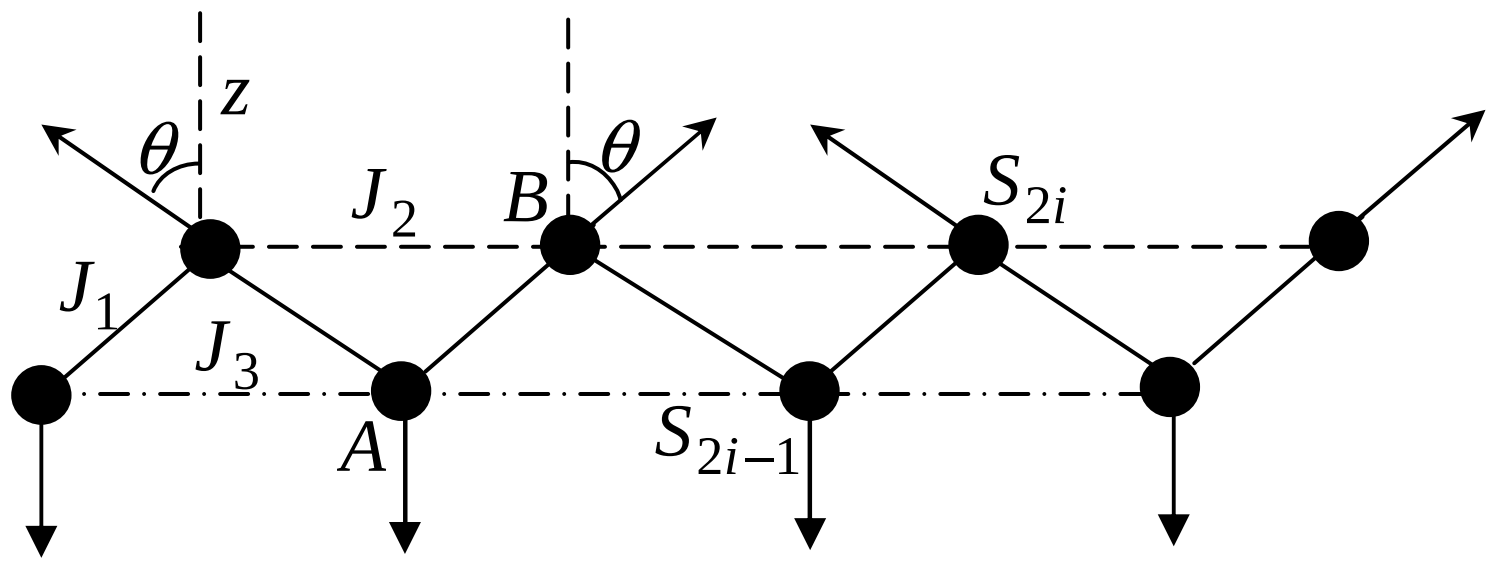

**Figure 2**

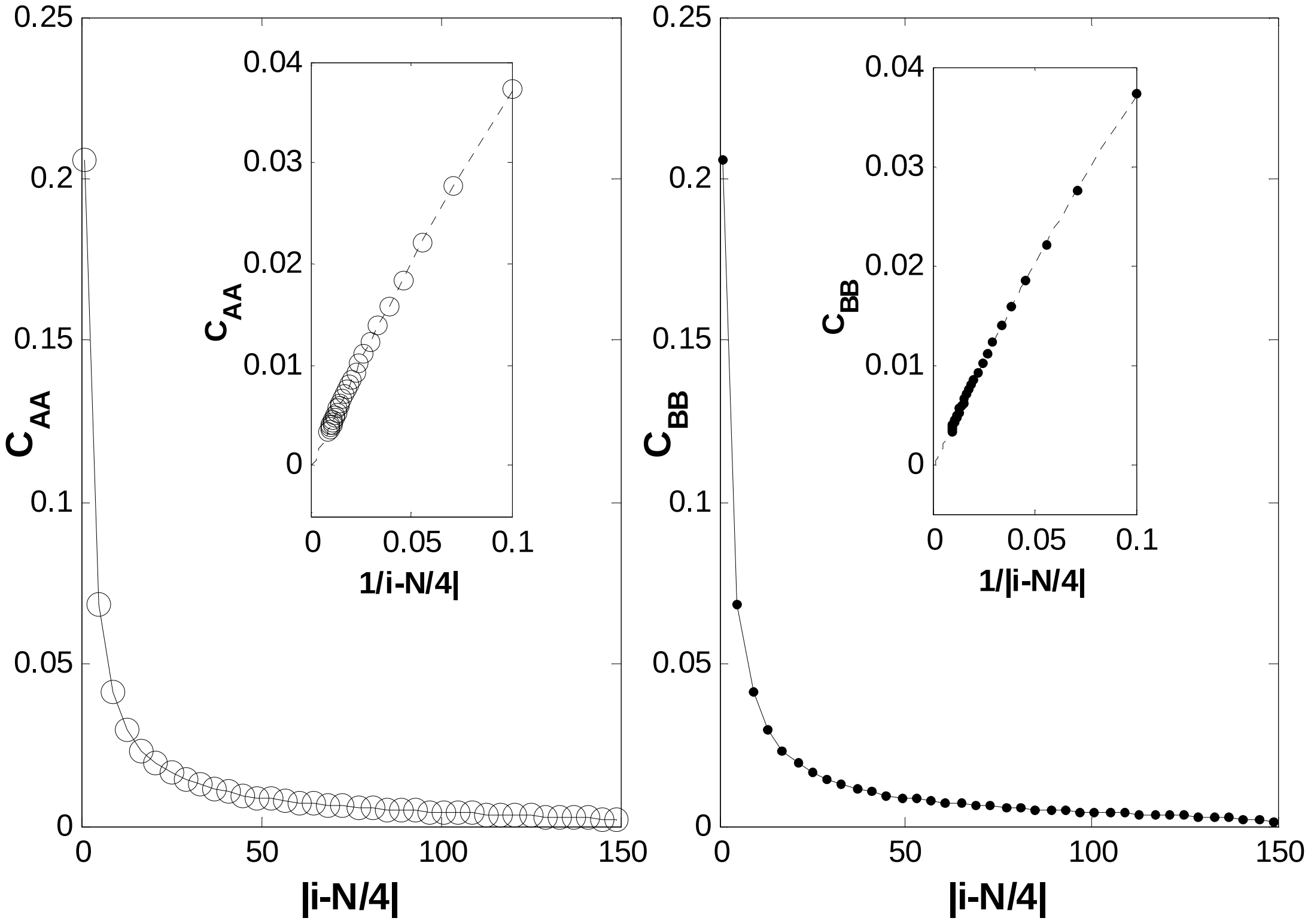
0.25
0.2
0.15
0.1
0.05
0
$C_{AA}$
0 50 100 150
|i-N/4|
0.04
0.03
0.02
0.01
0
$C_{AA}$
0 0.05 0.1
1/i-N/4|
0.25
0.2
0.15
0.1
0.05
0
$C_{BB}$
0 50 100 150
|i-N/4|
0.04
0.03
0.02
0.01
0
$C_{BB}$
0 0.05 0.1
1/|i-N/4|

**Figure 3**

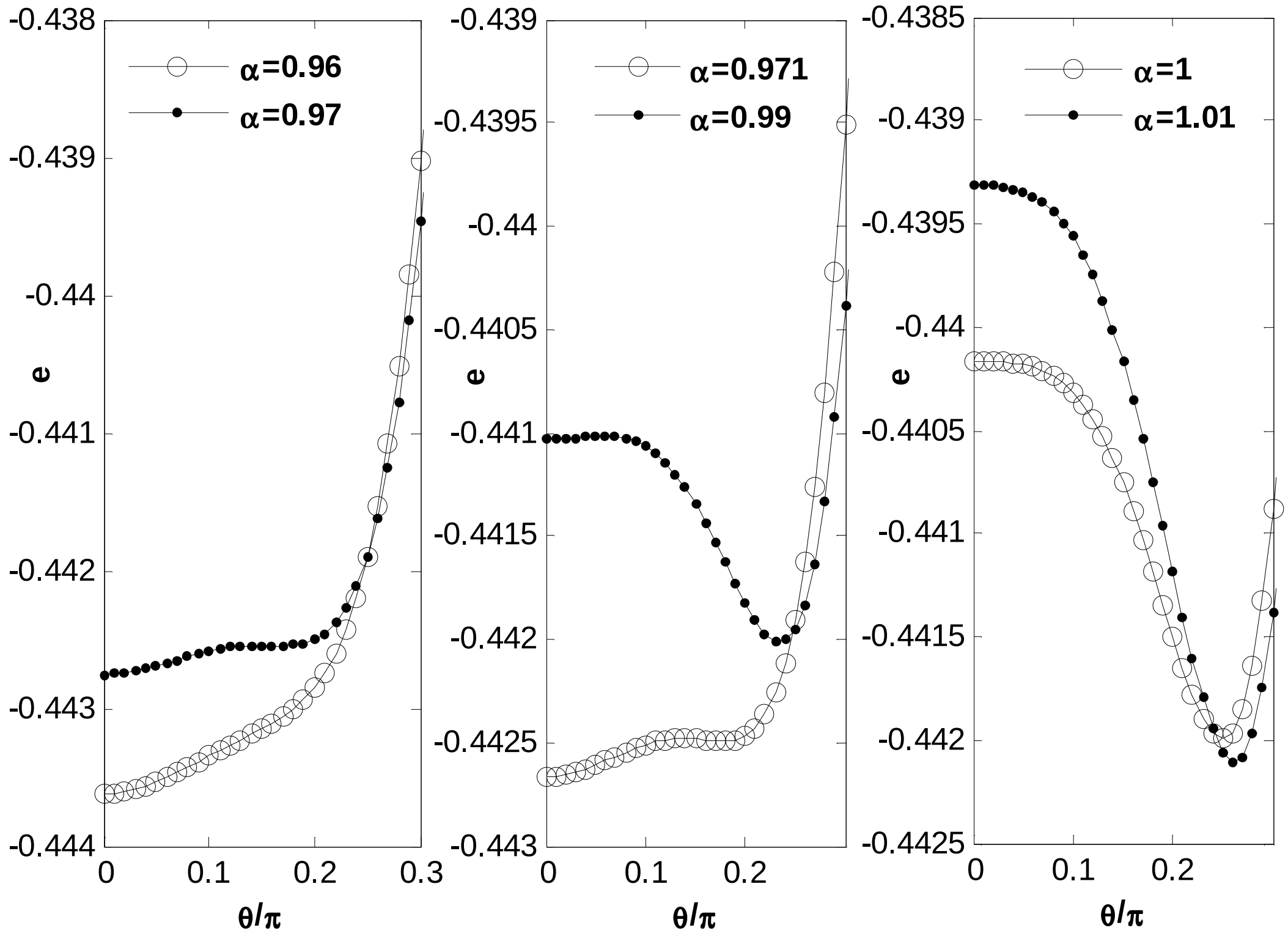

**Figure 4**

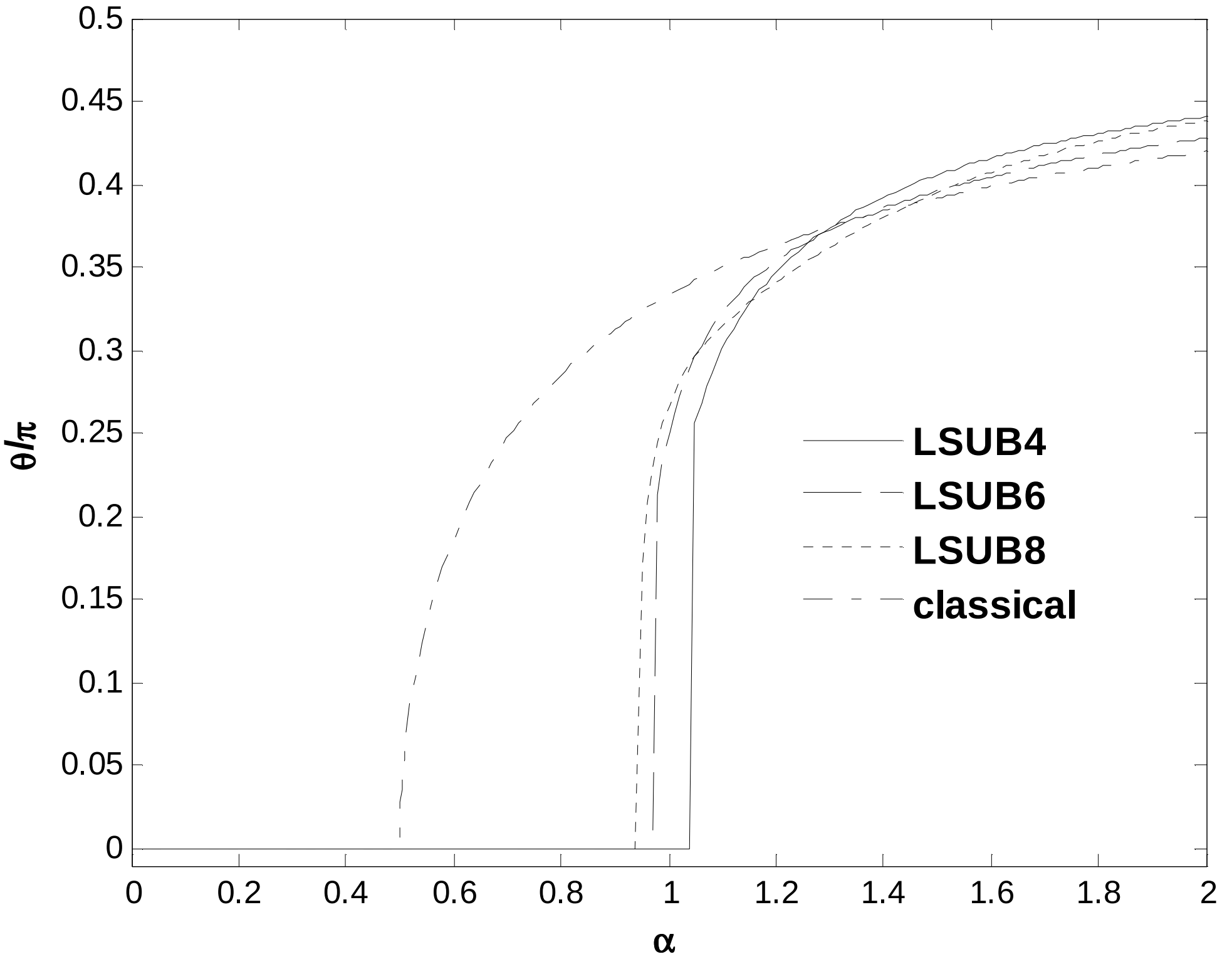

**Figure 5**

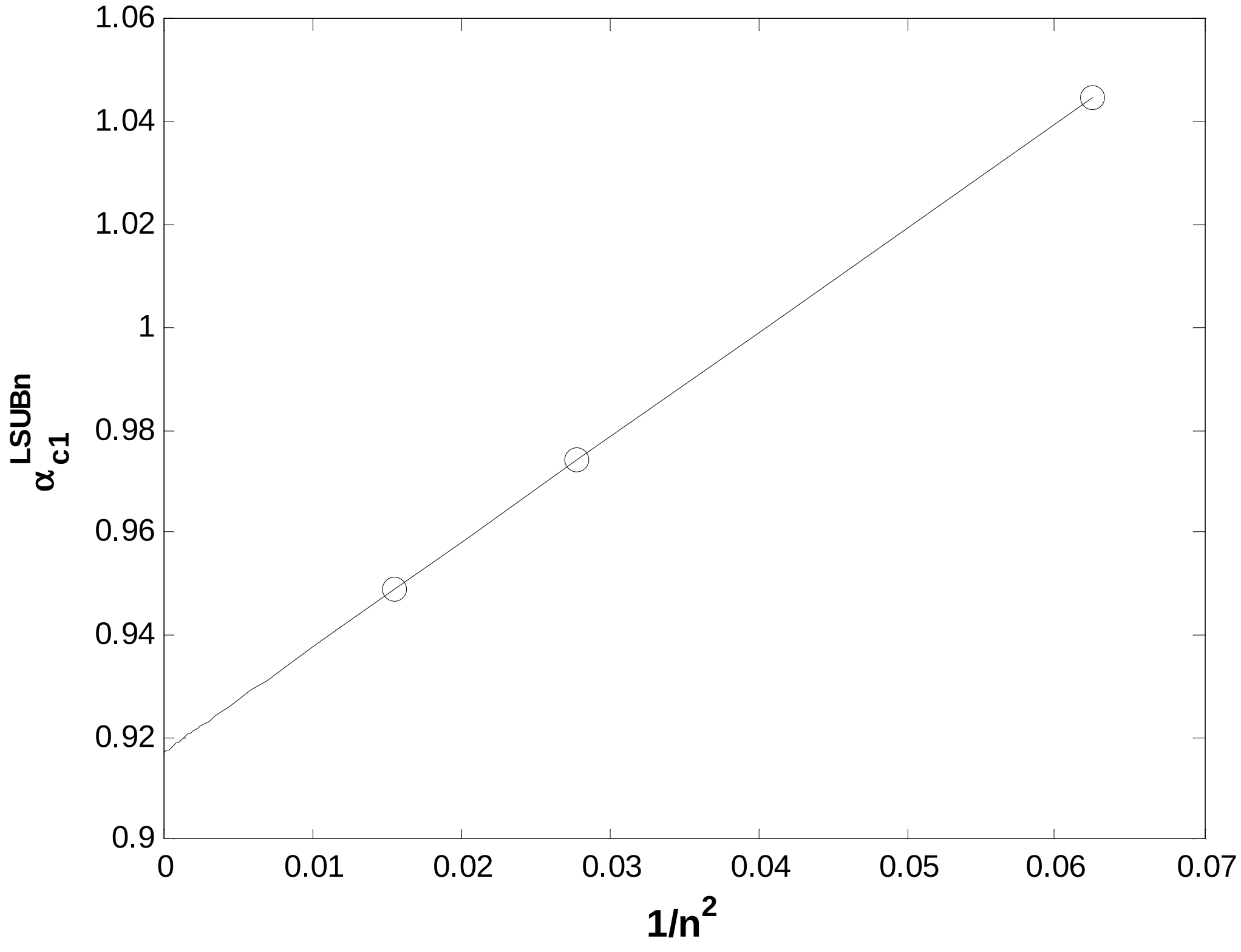

**Figure 6**

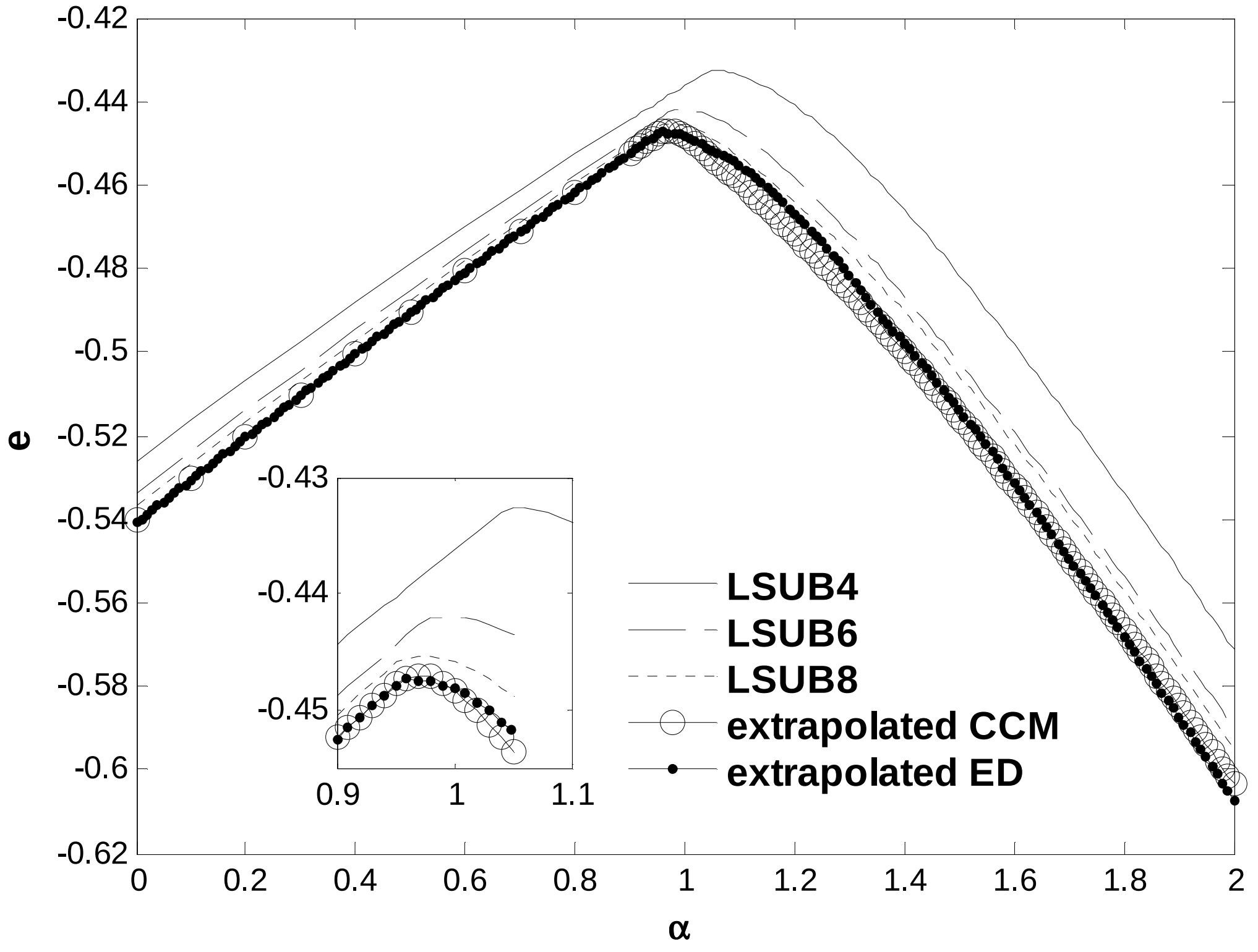

**Figure 7**

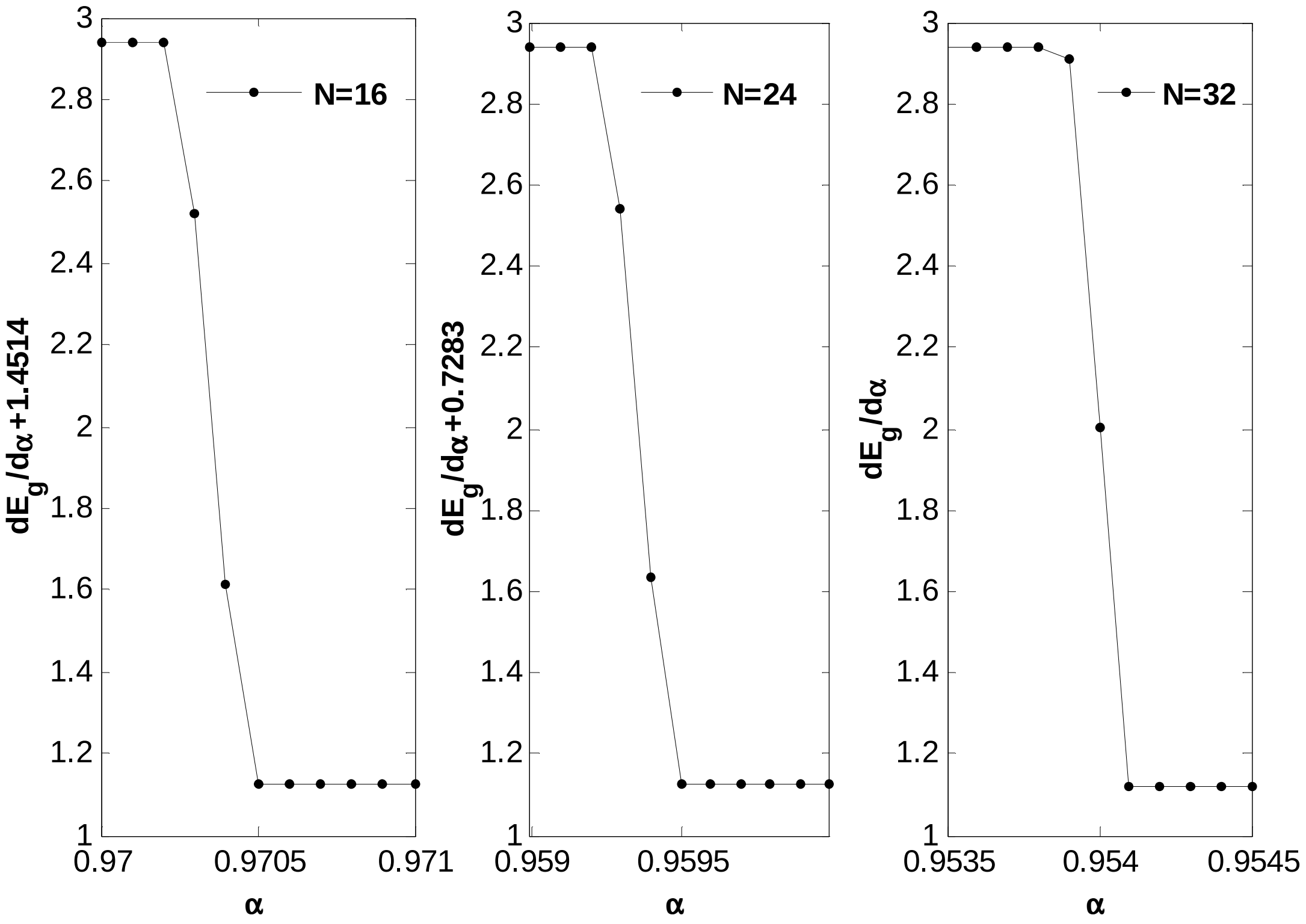

# Figure 8

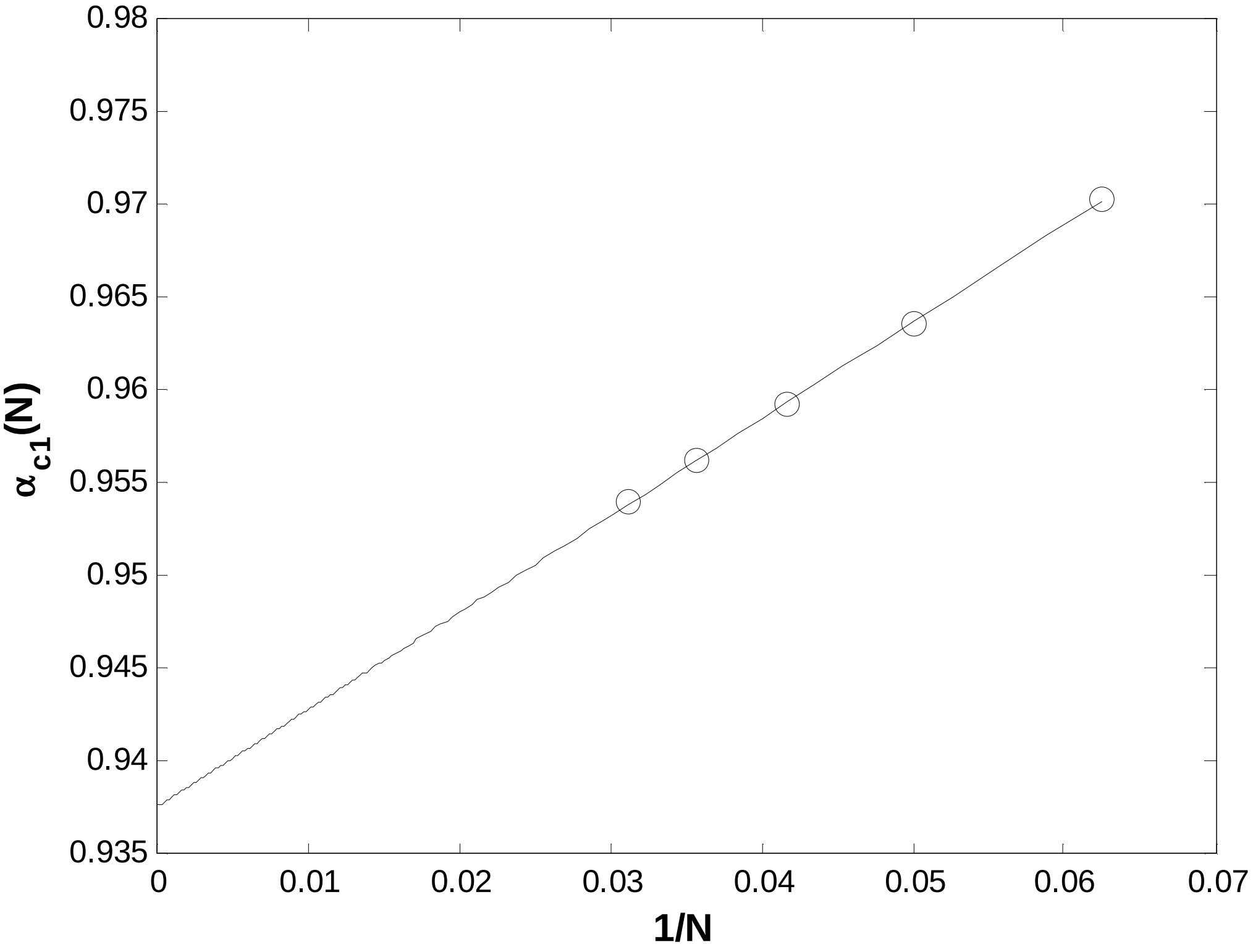

**Figure 9**

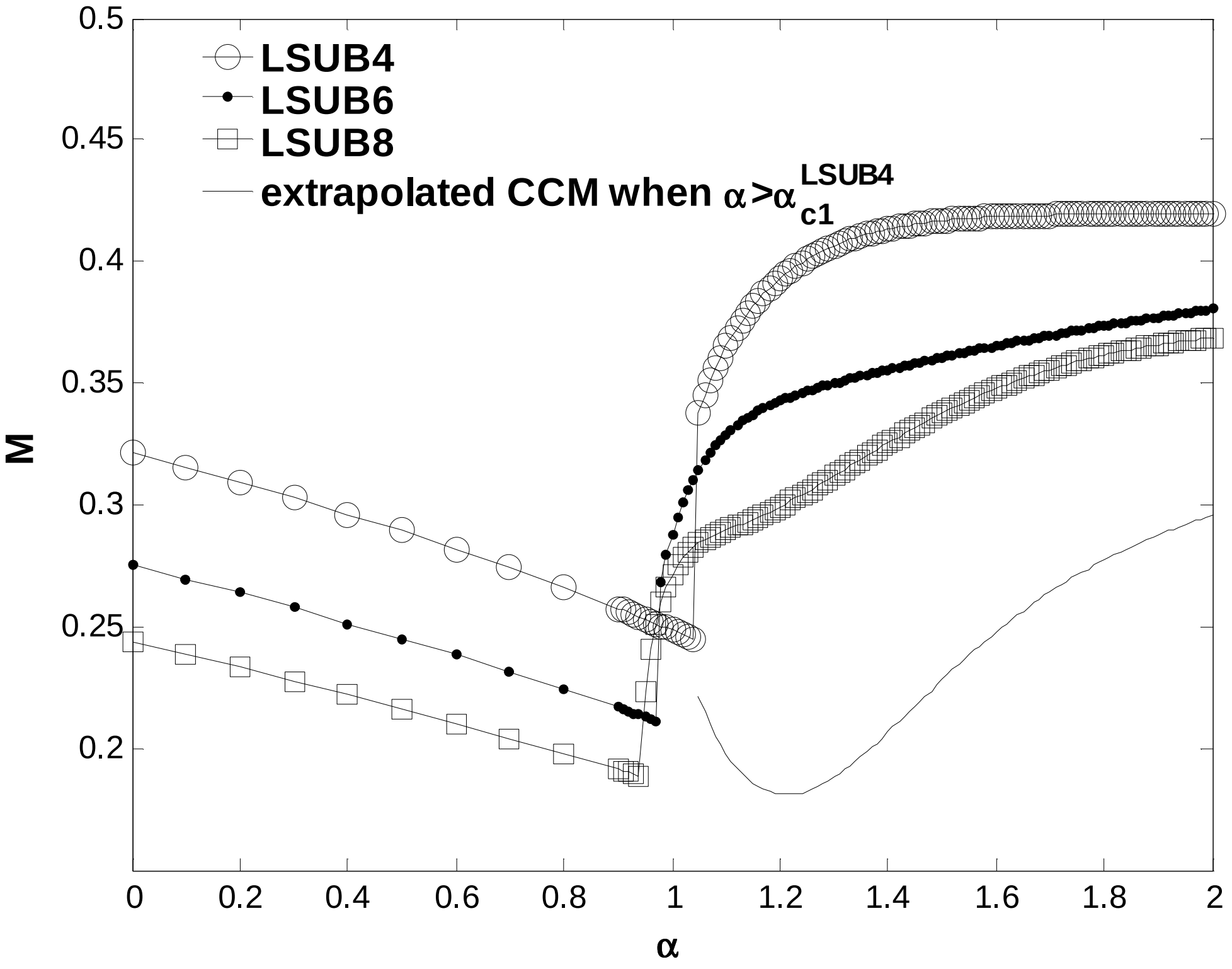

**Figure 10**

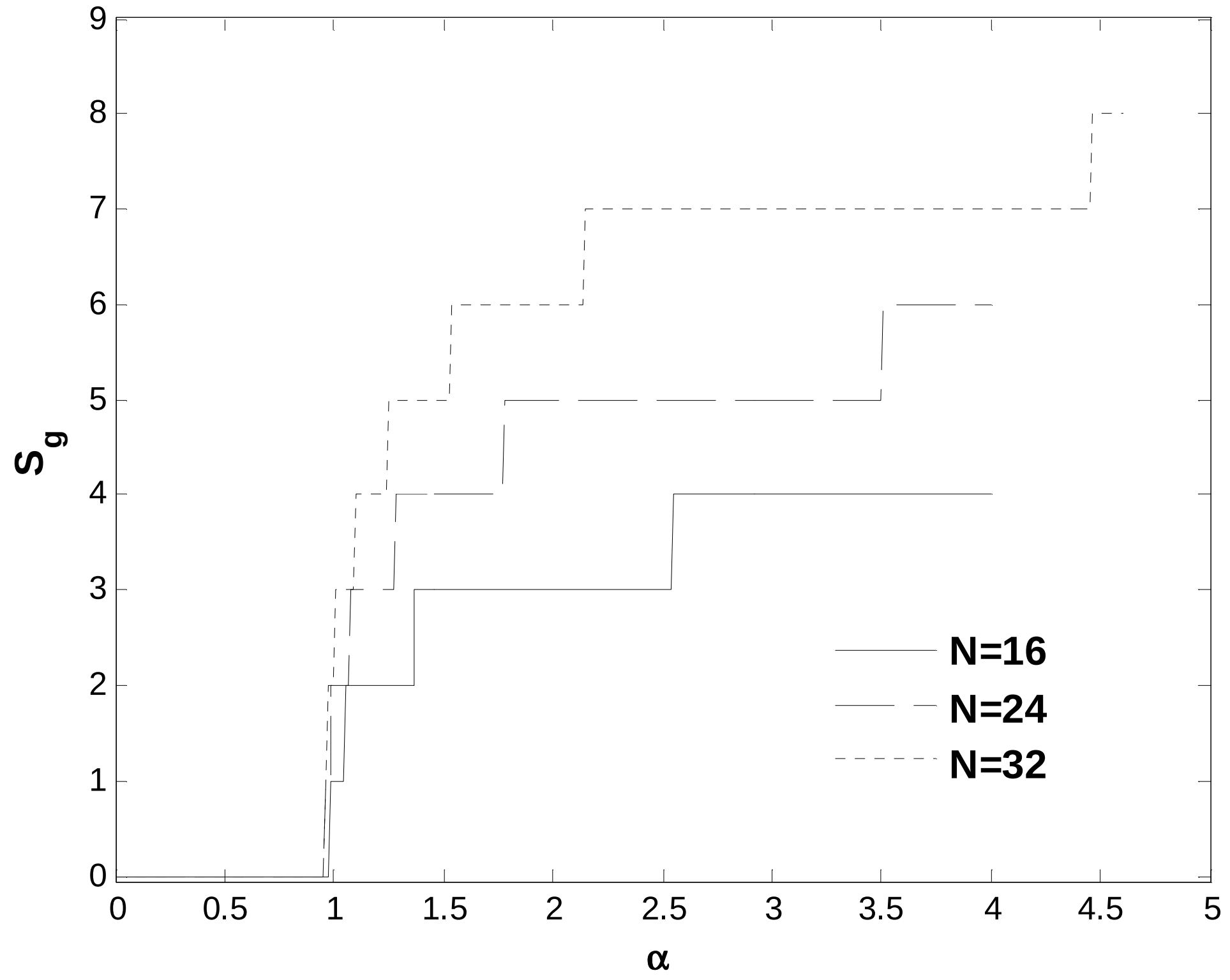

**Figure 11**

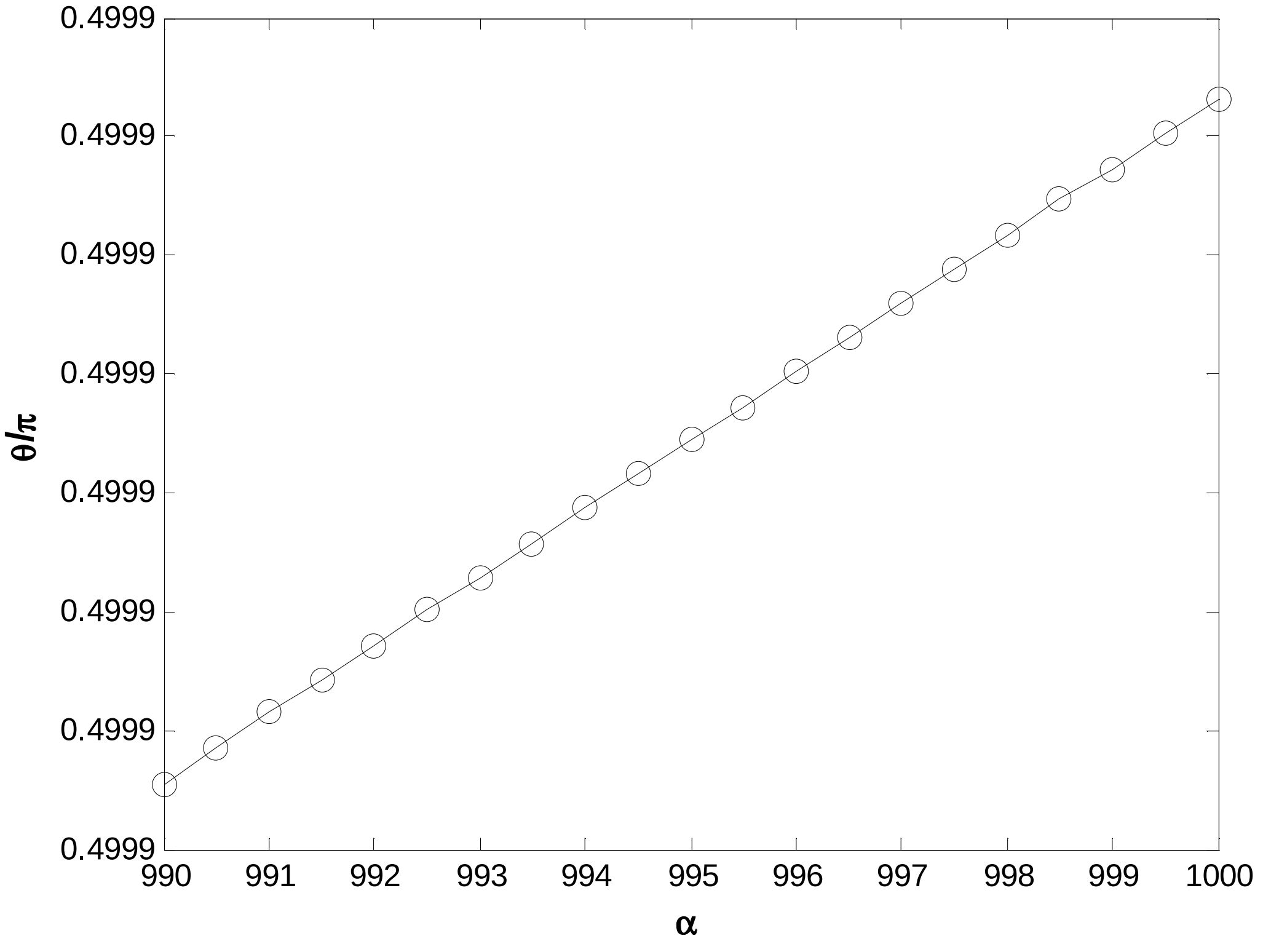

**Figure 12**

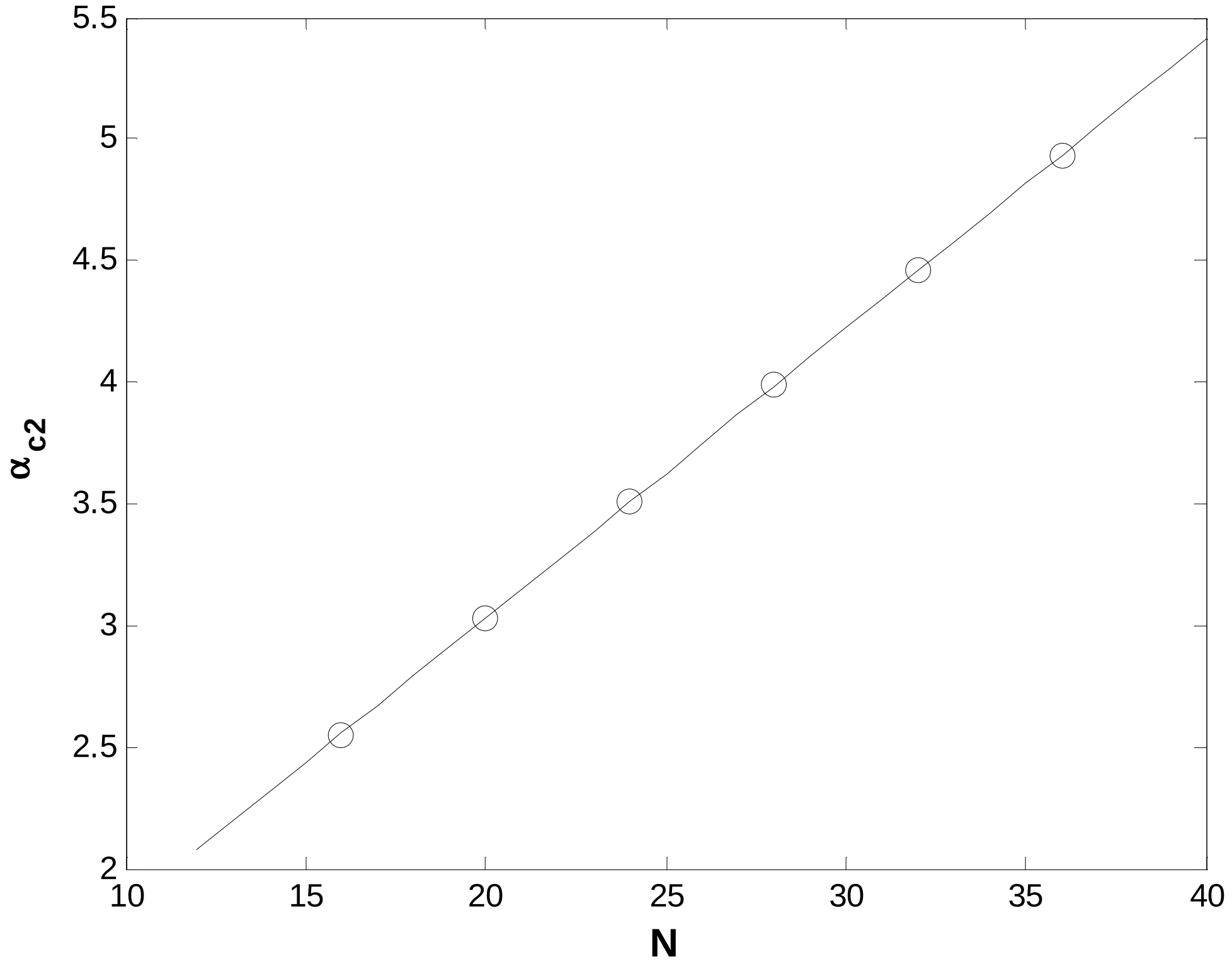

**Figure 13**

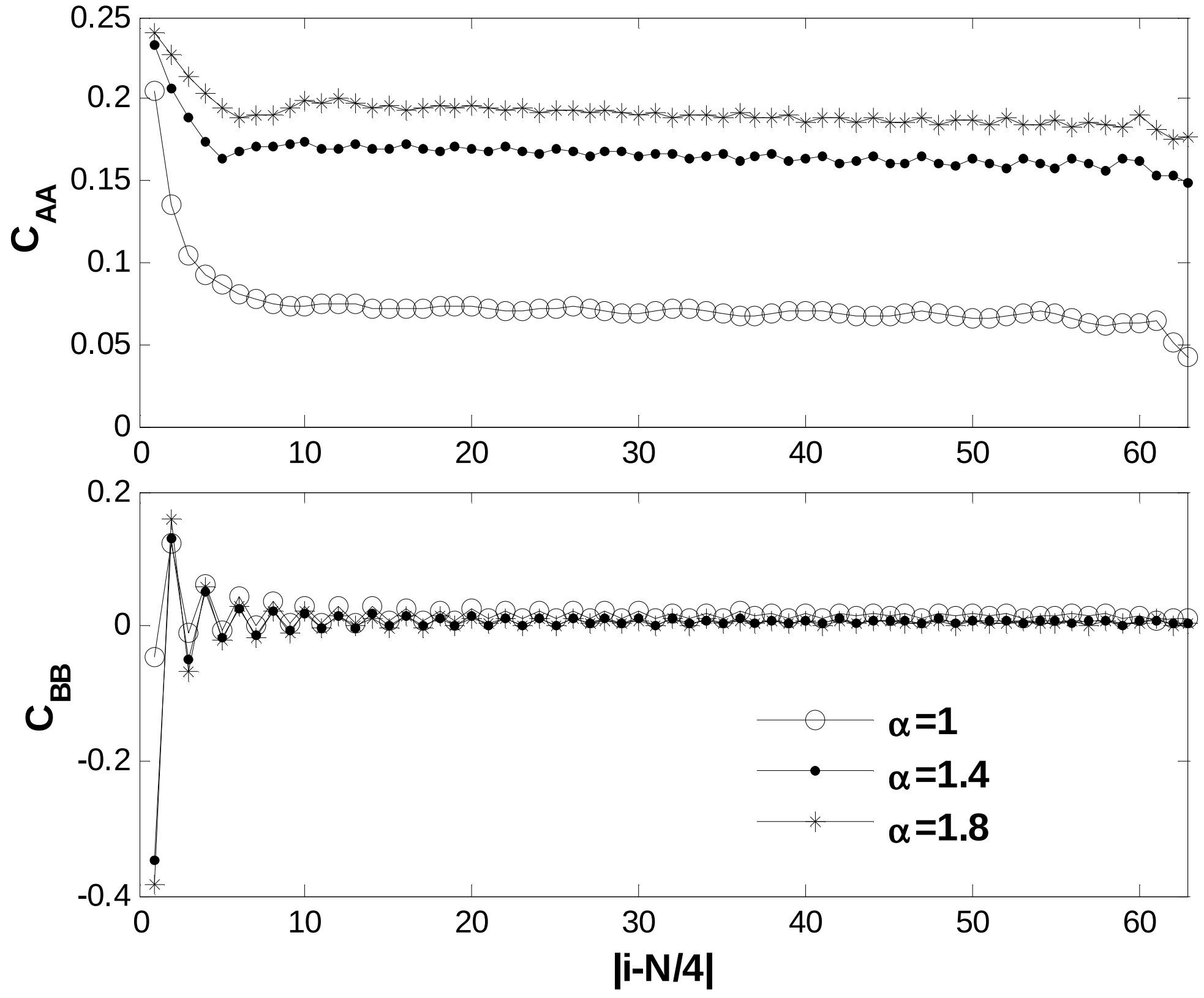

0.25
0.2
0.15
0.1
0.05
0
C_AA
0
10
20
30
40
50
60
0.2
0
-0.2
-0.4
C_BB
α=1
α=1.4
α=1.8
|i-N/4|